# Bilgi Güvenliği Kapsamında Yeni Bir Veri Şifreleme Algoritması Tasarımı ve Gerçekleştirilmesi

Nurettin TOPALOĞLU[1], M. Hanefi CALP[2], Burak TÜRK[3]

[1,3]Teknoloji Fakültesi, Bilgisayar Mühendisliği, Gazi Üniversitesi, Türkiye
[2]Yönetim Bilişim Sistemleri, Gazi Üniversitesi, Türkiye
nuretin@gazi.edu.tr, mhcalp@gazi.edu.tr, izciburakturk@gmail.com

***Özet*** – Günümüzde bilginin korunması, güvenliğinin sağlanması ve ihtiyaç halinde kayıpsız bir şekilde geri getirilmesi ciddi derecede önemli görülmektedir. Bilgi güvenliği, eldeki veriyi şifrelemek, korumak, iletmek ve paylaşmakla mümkündür. Bu noktada kriptoloji konusu gündeme gelmektedir. Şifre bilimi olan kriptoloji, herhangi bir metnin belli bir tekniğe göre şifrelenmesi ve bu mesajların güvenli bir şekilde alıcıya iletilmesi işlemlerini kapsamaktadır. Bu çalışmada, bilgi güvenliği kapsamında özgün bir veri şifreleme algoritması geliştirilmiştir. Geliştirilen algoritma, "Tekli Alfabetik Yer Değiştirme" tekniğine göre tasarlanmış olup algoritmanın oluşturulması sürecinde Sezar Şifreleme Tekniği, Çoklu Alfabeli Algoritmaları ve Enigma'dan da yararlanılmıştır. Çalışmanın amacı, yazılan bir metnin güvenli bir şekilde saklanması ve istenildiği takdirde deşifre edilerek bir bilgi güvenliğinin sağlanmasıdır. Yapılan çalışmada çözülmesi zor bir algoritma geliştirilmeye çalışılmış olup bu durum kullanılan şifre alfabe sayısı göz önüne alındığında açıkça ortaya çıkmaktadır. Çalışmanın bilgi güvenliği alanına katkı sağlaması beklenmektedir.

**Anahtar Kelimeler -** Kriptoloji, şifreleme, deşifreleme, bilgi güvenliği

# A Novel Data Encryption Algorithm Design and Implementation in Information Security Scope

**Abstract** – Today, the protection of information, ensuring of the safety and the recall in lossless in case of need is highly significant and it is seen as a major threat in the field. Information security is possible by hiding the available data, by encrypting, saving, transmitting and sharing. At this point, cryptology subject comes up. Cryptology called as password science includes with the encrypting according to a certain technique of different texts and the transmitting these messages to the recipient in a secure environment. In this study, in information security scope, an original data encryption algorithm was developed. Algorithm developed according to the "Single Characters Relocation" technique is designed. In the creation process of the algorithm, was also utilized from Caesar encryption algorithm, multi-alphabet algorithms and Enigma. The purpose of the study is to store the text written in a safe manner and thus to ensure information security is by deciphering upon request. In the study, an algorithm that is very difficult for its decryption was developed and this is clearly emerging when considering the number used passwords alphabet. This study is expected to contribute to the field of information security.

## 1. GİRİŞ (INTRODUCTION)

Günümüzde her alanda bilgisayar teknolojisinin yaygın olarak kullanılması ve özellikle bilgisayar ağlarının gelişmiş olması bilgiye erişimi kolaylaştırmaktadır. Bilgiye erişimin kolay olmasının yanında bilgi güvenliğini sağlamak da zaruri bir ihtiyaç olmaktadır. Dolayısıyla, bilginin tehditlere veya saldırılara karşı korunması ve güvenli bir şekilde transfer edilmesi önemli bir sorun haline gelmiştir [1].

İnternet kullanımının her geçen gün yaygınlaşmasıyla birlikte bilişim teknolojilerinde yer alan güvenlik zafiyeti veya açıklar da artmaya başlamıştır. Bilişim teknolojilerinde gizlilik, bütünlük ve sürekliliğin sağlanması amacıyla güvenlik konusuyla ilgili birçok ürün ve proje geliştirilmektedir. Bu durum, bilgi güvenliği konusunun önemini açıkça ortaya koymaktadır. Bilgi güvenliği, yetkisiz erişimlerden bilgiyi koruyarak sözkonusu bilginin gizliliğini (confidentiality), bütünlüğünü (entirety), doğruluğunu (integrity) ve erişilebilirliğini (availability) sağlamaktır. Bilgi güvenliği; teknoloji (yazılım ve donanım), insan, süreç, yöntem ve metodoloji gibi bir çok kavramı kapsamakta ve bilişim dünyası için oldukça önemli görülmektedir (Şekil 1)[2-4].

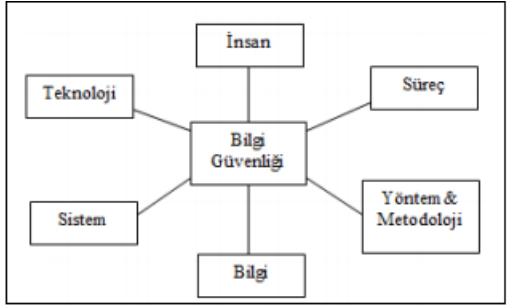

Şekil 1. Bilgi güvenliği kavramları
(Information security concepts)

Ayrıca bilgi güvenliği, "*Bilginin bir varlık olarak hasarlardan korunması, doğru teknolojinin, doğru amaçla ve doğru şekilde kullanılarak bilginin her türlü ortamda, istenmeyen kişiler tarafından ele geçirilmesini önlemek*" olarak tanımlanır [5,6]. Bu noktada, kriptografi konusuna dikkat çekmek gerekmektedir. Kriptografi, bir mesajın ya da bilginin geçici olarak okunamaz hale dönüştürülerek hedefine ulaştırılması ve karşı tarafta bilginin tekrar okunabilir hale dönüştürülmesi için kullanılan şifreleme konusuna verilmiş genel bir addır. Diğer bir tanımla kriptografi, güvensiz bir ortamda bilgi sızdırmadan iletişim sağlamak amacıyla matematiksel yöntemlerin geliştirilmesidir[1].

Ülkemizde şifreleme bilimi alanındaki çalışmalar 1995 yılında TÜBİTAK bünyesinde UEKAE (Ulusal Elektronik ve Kriptoloji Araştırma Enstitüsü) biriminin kurulmasıyla başlamıştır. Daha sonra ODTÜ'de Kriptografi Bölümü açılmıştır. Ülkemiz bu alandaki çalışmalarına her ne kadar diğer ülkelerden daha sonra başlamış olsa da yapılan çalışmalarla önemli bir yer edinmiştir [7]. Günümüzde ise şifreleme işlemleri, özel veya devlet kurumları olmak üzere hemen her alanda kullanılmaktadır. Film ve kitaplardaki olaylar içerisinde dahi sözkonusu gizliliği görmek mümkündür. Bunların çözümünde şifrelenmiş mesajların kullanılması ve her şifreleme tekniğinin farklı mantıklara dayanması kişilerde bu alana yönelik ilgi ve merak uyandırmaktadır.

Bu çalışmada gelişmesini sürdürmekte olan ve gelecekte daha da önemli bir alan haline gelmesi beklenen kriptografiye yeni bir bakış açısıyla yeni bir teknik kullanılarak özgün bir algoritma geliştirilmiştir. Çalışmada ikinci bölümde, kriptoloji algoritmaları ve kullanımları; üçüncü bölümde, geliştirilen Veri Şifreleme Algoritması'nın ayrıntıları; son olarak 4. bölümde ise, çalışmanın sonuç ve önerileri açıkça ortaya konmuştur.

## 2. KRİPTOLOJİ ALGORİTMALARI VE KULLANIMLARI (CRYPTOGRAPHY ALGORITHMS AND USAGES)

Kriptoloji, bir şifre bilimi olup, çeşitli mesajların belli bir yöntemle şifrelenmesi, bu mesajların güvenli bir ortamda alıcıya iletilmesi ve yeniden çözülmesiyle (deşifre edilmesi) ilgilidir [8]. Şifreleme özellikle günümüzde bilgilerin korunması, saklanması, iletilmesi ve değerlendirilmesi bakımından oldukça önemlidir. Şifreleme sistemleri tarihin akışında önemli değişiklikler meydana getirmiştir. Çözülen ya da çözülemeyen şifreli metinler büyük yıkımlara neden olmuşlardır. Bilgi herkes için çok önemli bir değer olup günümüze kadar geçen süreçte, toplumların sürekliliğini sağlamak için korunması gereken bir güç olarak algılanmıştır [9]. Şifreleme için tarih boyunca çok çeşitli yöntemler denenmiştir [10]. Sözkonusu yöntemlerin birkaçı özellikleriyle kısaca şöyledir:

*2.1. Atbash Şifreleme Sistemi (Atbash Encryption System)*

İbrani peygamber Yeremya'nın kehanet ve şifrelerini yazarken kullanmış olduğu şifreleme sistemidir. Sistemin kullanımı şu şekildedir: İbranice alfabesinde yer alan ilk harf, son harfle, ikinci harf sondan bir önceki harfle, üçüncü harf ise sondan iki önceki harf ile yer değiştirmektedir [7]. ATBASH şifresi Türkçe'ye uyarlanarak örnek verilirse, şifreleme için gerekli olan tablo Şekil 2'de görüldüğü gibi olacaktır.

| ALFABE: | A | B | C | Ç | D | E | F | G | Ğ | H | I | İ | J | K | L | M | N | O | Ö | P | R | S | Ş | T | U | Ü | V | Y | Z |
|---|---|---|---|---|---|---|---|---|---|---|---|---|---|---|---|---|---|---|---|---|---|---|---|---|---|---|---|---|---|
| ŞİFRE : | Z | Y | V | Ü | U | T | Ş | S | R | P | Ö | O | N | M | L | K | J | İ | I | H | Ğ | G | F | E | D | Ç | C | B | A |

Şekil 2. ATBASH şifreleme sistemi için kullanılan alfabe (The alphabet used for ATBASH encryption system)

<u>Örnek cümle:</u>   Bugün meydanda buluşacağız.
<u>Şifrelenmiş Metin:</u> ydscj ktbuzjuz ydldfzvzröa

*2.2. Ebced Hesabı (Ebced Acoount)*

Hemen her dilin alfabesinde yer alan harflerin rakamsal olarak karşılığı olduğu bilinmektedir. Bunlar arasında en çok bilineni ise İbrani-Süryani, Grek ve Latin harf-sayı sistemleridir. "Ebced Hesabı" olarak kullanılan, Arap alfabesinin ebcet düzenine dayanan rakamlar ve hesap sistemi günümüzde de halen kullanılmaktadır [11]. Ebced hesabındaki temel mantık; alfabetik bir sayı sistemini kullanarak kelimelerin sayısal değerlerinin hesaplanmasına dayanmaktadır [12].

*2.3. Skytale Tekniği* (Scytal Technique)

Tarihteki ilk şifreleme kalıntılarının sivil amaçlarla olduğu görülmektedir. Askeri amaçlı olarak ilk kullanımı, Yunanlılar tarafından M.Ö. 5. yüzyılın başlarında "Skytale" adını verdikleri şifreleme cihazıyla olmuştur. Skytale aynı zamanda kullanılan ilk kriptografik cihazdır. Skytale kullanılarak bir mesajı şifrelemek için öncelikle uzun bir parşömen ya da papirüs, silindirik bir sopa etrafına sarılmaktadır (Şekil 3). Gizlenecek mesaj, uzunlamasına papirüs sarılı sopa üzerine, her bir şerit turunda bir harf gelecek şekilde yazılmaktadır. Şerit açılıp kaldırıldıktan sonra anlamsız

*2.4. Steganografi* (Steganography)

Steganografi tekniği, eski Yunancada "gizlenmiş yazı" anlamına gelmektedir. Dijital ortamda bulunan ses, fotoğraf ve video gibi büyük boyutlardaki dosyalara bir verinin gizlenmesiyle anlaşılmayacak şekilde iletilmesi sağlanmaktadır [6,10].

*2.5. Polybius Şifrelemesi* (Polybius Encryption)

Polybius dama tahtası, alfabenin harflerini içeren beşe beşlik bir ızgaradan oluşmaktadır. Her harf, ilki harfin bulunduğu satır ve ikincisi de sütun olmak üzere iki sayıya dönüştürülmektedir. A harfi için 11, B harfi için 12 ve sonraki harfler için ise ilgili sayılar eşleşmektedir. Burada dikkat edilmesi gereken konu, ilk rakamın satırı ikinci rakamın sütunu göstermesidir. Polybius şifreleme tekniğinde kullanılan ızgara 5x5 birimlik değil, kullanılan alfabe hatta şifrelenecek metnin içerdiği alfabe harfleri sayısına göre istenilen boyutlarda (nxm) tasarlanabilmektedir [7]. Şekil 4'te verilen Polybius'un şifresinin Türkçe alfabe kullanılarak oluşturulan örneğinde harf sayısı sebebiyle 5x6'lık bir matris kullanılmıştır.

|   | 1 | 2 | 3 | 4 | 5 | 6 |
|---|---|---|---|---|---|---|
| 1 | A | B | C | Ç | D | E |
| 2 | F | G | Ğ | H | I | İ |
| 3 | J | K | L | M | N | O |
| 4 | Ö | P | R | S | Ş | T |
| 5 | U | Ü | V | Y | Z |   |

Şekil 4. Polybius'un dama tahtası (Türkçe harflerle)
(Polybius checkerboard (Turkish letters))

Örnek Cümle: Gazi Üniversitesi

Örneğin;
Açık Metin: Gazi Üniversitesi Teknik Eğitim Fakültesi
Satır Sayısı: 2

Harflerin oluşturduğu şifreli metin elde edilmektedir. Deşifreleme işlemi için, şifrelemede kullanılan sopa ile aynı çapta ve uzunlukta bir sopaya sahip olmak gerekmektedir. Sopanın çapındaki en ufak bir farklılık anlamsız metinlerin ortaya çıkmasına sebep olmaktadır [11].

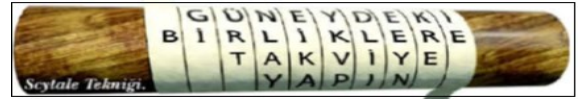
Şekil 3. Skytale tekniği (Scytal technique)

Şifreli Cümle: 22-11-55-26-52-35-26-53-16-43-44-26-46-16-44-26

*2.6. Zigzag Şifrelemesi* (Zigzag Encryption)

Zigzag Şifrelemesi var olan karakterlerin belirli bir permütasyon kullanılarak yer değiştirmesi prensibine dayanan bir yöntemdir. Adını şifreleme yönteminin biçiminden dolayı bu şekilde almıştır. Şifrelenecek olan metin aşağı doğru ve diagonal olarak hayali raylara doğru iner. En alttaki hayali raya geldiğinde tekrar yukarı doğru çıkar ve en üste ulaştığında tekrar aşağı doğru inerek açık metnin uzunluğuna göre devam eder (Şekil 5) [13].

Şifrelenmiş Metin:
GZÜİESTSTKİEİİFKLEİAİNVRİEİENKĞTMAÜTS

Şifreli Metnin Çözümü: Şifreli mesajın açılması için satır sayısı bilinmelidir. Satır sayısı anahtar olarak iki taraf arasında daha önceden kararlaştırılan bir sayıdır.

Gelen şifreli metindeki harflerin sayısı çift veya tek olma durumu bulunmaktadır. Şayet sayı çift ise, şifreli metin ikiye bölünerek harfler sırayla bir ilk yarımdan bir de ikinci yarımdan alınmaktadır. Bu yolla düz metin elde edilmektedir. Şayet sayı tek ise, yukarıda verilen örnekteki gibi ilk satır ikinciden bir fazla olacaktır. Bu sebeple, şifreli metindeki harflerin sayısının bir eksiği alınarak elde edilen sayı ikiye bölünmektedir. Elde edilen sonucun bir fazlası algoritmanın ilk satırında yer alan harf sayısını, çıkan sonuç da ikinci satırındaki harf sayısını vermektedir. "Gazi Üniversitesi Teknik Eğitim Fakültesi" örneğindeki şifre 37 harf içermektedir. Bunun ilk 19 harfi ilk satırı, geriye kalan 18 harfi de ikinci satırı vermelidir.

Bu şekilde şifreli metin ayrılırsa:
GZÜİESTSTKİEİİFKLEİ/AİNVRİEİENKĞTMAÜTS
elde edilir. Harfler sırasıyla bir ilk kısımdan bir de ikinci kısımdan alınırsa düz metin elde edilmiş olunur.

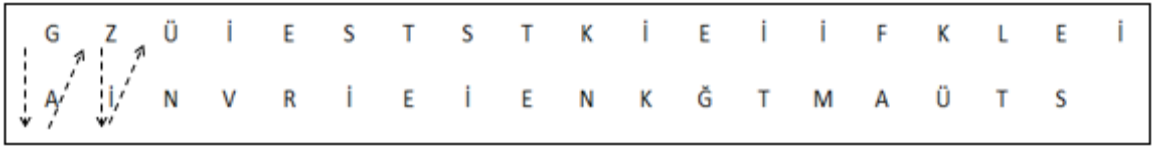

Şekil 5. Açık metnin zigzag biçimde gösterimi (Zigzag format display of open text)

*2.7. Sezar Şifreleme Yöntemi (Caesar Encryption Method)*

"Sezar Şifrelemesi", alfabedeki her bir harfin belli sayıda ileriye kaydırılmasına dayalı bir şifreleme metodudur [7].

Örneğin,

| ALFABE : | A | B | C | Ç | D | E | F | G | Ğ | H | I | İ | J | K | L | M | N | O | Ö | P | R | S | Ş | T | U | Ü | V | Y | Z |
|---|---|---|---|---|---|---|---|---|---|---|---|---|---|---|---|---|---|---|---|---|---|---|---|---|---|---|---|---|---|
| ŞİFRE : | Ç | D | E | F | G | Ğ | H | I | İ | J | K | L | M | N | O | Ö | P | R | S | Ş | T | U | Ü | V | Y | Z | A | B | C |

Şekil 6. Sezar şifreleme yöntemi (Caesar encryption method)

Örnek Cümle: Gazi Üniversitesi
Şifreli Cümle: IÇCL ZPLATULVĞUL

Örnekte verilen şifre kullanılarak mesaj gönderen şahıs, mesajda yer alan her harfin yerine şifre alfabedeki karşılığını koyarak şifre metni elde eder veya tablo oluşturmadan mesajdaki her harfi üç harf sola kaydırır. Alan şahıs ise aynı yöntemle tablo oluşturarak şifrelenmiş metindeki her harfin üstündeki harfi alarak veya her harfi 26 harf sola kaydırarak mesajını çözmektedir.

*2.8. Alberti Diski (Alberti Disk)*

1404-1472 yılları arasında yaşamış olan Leon Alberti, 1466-1467 yılları arasında ilk kez çoklu alfabe kullanarak kriptolama yapmıştır. Bu yöntemde, Sezar şifreleme yöntemine benzer olarak harf kaydırma tekniği uygulanmaktadır. Sezar şifrelemeden farklı olarak kaydırma miktarı sabit olmayıp, kullanıcının kararına göre belirlenmektedir. Kriptolanacak metinde her harfin kriptolu karşılığı, Alberti Diski yardımıyla bulunmaktadır. İçteki çemberi sabit, dıştaki çemberi onun etrafında dönebilen bu disk yardımıyla, her harfin istenilen miktarda ötelenmiş hali kolaylıkla görülebilmektedir (Şekil 7) [7].

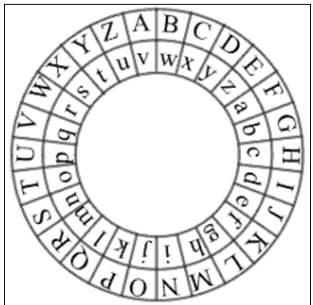

Şekil 7. Alberti Diski (Alberti Disk)

*2.9. Vigenere Şifresi (Vigenere Password)*

Bu şifreleme yönteminde, düz metinde bulunan her bir harf farklı bir şifre alfabeyle şifrelenmektedir. Hangi şifre alfabenin belirleneceği anahtar sözcüğe bakılarak karar verilmektedir. Böylece düz metindeki aynı sözcükler için farklı şifre metinler oluşturulmuş olunur [14].

Metin: Taarruz Dokuzda
Anahtar: Kale

| T | A | A | R | R | U | Z | D | O | K | U | Z | D | A |
|---|---|---|---|---|---|---|---|---|---|---|---|---|---|
| K | A | L | E | K | A | L | E | K | A | L | E | K | A |

Metnin altına, sürekli kendini tekrar edecek şekilde anahtarı yerleştirip çizelgeden eşleşmeler alınır. Vigenere Tablosundan eşleştirmeler yapılarak şifre metin elde edilir (Şekil 8).

Şifre Metin: DALVBUKHYKFDNA

|   | A | B | C | D | E | F | G | H | I | J | K | L | M | N | O | P | Q | R | S | T | U | V | W | X | Y | Z |
|---|---|---|---|---|---|---|---|---|---|---|---|---|---|---|---|---|---|---|---|---|---|---|---|---|---|---|
| A | A | B | C | D | E | F | G | H | I | J | K | L | M | N | O | P | Q | R | S | T | U | V | W | X | Y | Z |
| B | B | C | D | E | F | G | H | I | J | K | L | M | N | O | P | Q | R | S | T | U | V | W | X | Y | Z | A |
| C | C | D | E | F | G | H | I | J | K | L | M | N | O | P | Q | R | S | T | U | V | W | X | Y | Z | A | B |
| D | D | E | F | G | H | I | J | K | L | M | N | O | P | Q | R | S | T | U | V | W | X | Y | Z | A | B | C |
| E | E | F | G | H | I | J | K | L | M | N | O | P | Q | R | S | T | U | V | W | X | Y | Z | A | B | C | D |
| F | F | G | H | I | J | K | L | M | N | O | P | Q | R | S | T | U | V | W | X | Y | Z | A | B | C | D | E |
| G | G | H | I | J | K | L | M | N | O | P | Q | R | S | T | U | V | W | X | Y | Z | A | B | C | D | E | F |
| H | H | I | J | K | L | M | N | O | P | Q | R | S | T | U | V | W | X | Y | Z | A | B | C | D | E | F | G |
| I | I | J | K | L | M | N | O | P | Q | R | S | T | U | V | W | X | Y | Z | A | B | C | D | E | F | G | H |
| J | J | K | L | M | N | O | P | Q | R | S | T | U | V | W | X | Y | Z | A | B | C | D | E | F | G | H | I |
| K | K | L | M | N | O | P | Q | R | S | T | U | V | W | X | Y | Z | A | B | C | D | E | F | G | H | I | J |
| L | L | M | N | O | P | Q | R | S | T | U | V | W | X | Y | Z | A | B | C | D | E | F | G | H | I | J | K |
| M | M | N | O | P | Q | R | S | T | U | V | W | X | Y | Z | A | B | C | D | E | F | G | H | I | J | K | L |
| N | N | O | P | Q | R | S | T | U | V | W | X | Y | Z | A | B | C | D | E | F | G | H | I | J | K | L | M |
| O | O | P | Q | R | S | T | U | V | W | X | Y | Z | A | B | C | D | E | F | G | H | I | J | K | L | M | N |
| P | P | Q | R | S | T | U | V | W | X | Y | Z | A | B | C | D | E | F | G | H | I | J | K | L | M | N | O |
| Q | Q | R | S | T | U | V | W | X | Y | Z | A | B | C | D | E | F | G | H | I | J | K | L | M | N | O | P |
| R | R | S | T | U | V | W | X | Y | Z | A | B | C | D | E | F | G | H | I | J | K | L | M | N | O | P | Q |
| S | S | T | U | V | W | X | Y | Z | A | B | C | D | E | F | G | H | I | J | K | L | M | N | O | P | Q | R |
| T | T | U | V | W | X | Y | Z | A | B | C | D | E | F | G | H | I | J | K | L | M | N | O | P | Q | R | S |
| U | U | V | W | X | Y | Z | A | B | C | D | E | F | G | H | I | J | K | L | M | N | O | P | Q | R | S | T |
| V | V | W | X | Y | Z | A | B | C | D | E | F | G | H | I | J | K | L | M | N | O | P | Q | R | S | T | U |
| W | W | X | Y | Z | A | B | C | D | E | F | G | H | I | J | K | L | M | N | O | P | Q | R | S | T | U | V |
| X | X | Y | Z | A | B | C | D | E | F | G | H | I | J | K | L | M | N | O | P | Q | R | S | T | U | V | W |
| Y | Y | Z | A | B | C | D | E | F | G | H | I | J | K | L | M | N | O | P | Q | R | S | T | U | V | W | X |
| Z | Z | A | B | C | D | E | F | G | H | I | J | K | L | M | N | O | P | Q | R | S | T | U | V | W | X | Y |

Şekil 8. Vigenere şifre tablosu (Vigenere code table)

*2.10. Jefferson Diski (Jefferson Disk)*

Jefferson Diski, İngiliz alfabesinin harf sayısı nedeniyle 26 diskten oluşmaktadır. Her disk üzerinde alfabedeki tüm harfler rastgele yazılmıştır. Şifreleme işlemini yapmak isteyen kişi, düz metni Jefferson Diski'ndeki bu sırada oluşturup geri kalan sıralardan herhangi birini seçerek şifre metnini elde etmektedir. Aynı özellikleri taşıyan Jefferson Diski, hem alıcı hem de göndericide bulunmak zorundadır. Şifre metninin elde eden alıcı, elindeki Jefferson Diski ile şifre metnini oluşturup, geri kalan sıralardaki anlamlı metni çıkarmaktadır [11].

*2.11. Whatsone-Palyfair Şifresi (Whatsone-Palyfair Password)*

Charles Wheatsone ve Baron Lyon Palyfair tarafından geliştirilen bu teknikte, 5x5 bir matris kullanılmaktadır. Aslında bu şifreleme sistemi 26 harften oluşan İngiliz alfabesi için tasarlanmış olup "I" ve "J" harfleri beraber düşünülmüştür. 25 hücresi olan matrise tüm alfabe her hücreye bir harf gelecek şekilde sığmaktadır. Palyfair şifresinde şifreleme algoritması, düz metindeki her bir harf çiftini başka bir harf çiftiyle değiştirmektedir; fakat şifreleme süreci başlatılmadan önce bir anahtarın belirlenmesi ve belirlenen bu anahtara göre matrisin oluşturulması gerekmektedir. Anahtar, matrisin ilk hücresinden başlayarak, her hücreye bir harf yazılacak şekilde birinci satırdan başlanır. Birden fazla kullanılan harfler ilk kullanımdan sonra atılır ve ilk satıra sığmadığı takdirde ikinci satırdan devam edilir. Anahtar yazılıp, kalan hücrelere de alfabenin geri kalan harfleri sırayla yazılır. Anahtar sözcüğü kriptografi olarak seçilir ve birden fazla kullanılan harfleri atarak anahtar sözcüğü kriptograf olarak ilk satırdan itibaren yazılır. Örneğin,

Anahtar : KRİPTOGAF

Herşeyden önce şifrelenecek mesaj, harf çiftlerine ayrılır. Şayet bu mesaj tek sayıda harften meydana geliyorsa, mesajın sonuna istenilen bir harf eklenmektedir. Deşifreleme işleminde son harfin herhangi bir önemi bulunmamaktadır. Her ikilideki harfler birbirinden farklı olmalıdır.

Tablo 1. Anahtar kelimeye göre hazırlanan tablo (The table prepared according to keyword)

| K | R | İ | P | T |
|---|---|---|---|---|
| O | G | A | F | B |
| C | Ç | D | E | Ğ |
| H | I | J | L | M |
| N | Ö | S/Ş | U/Ü | V/Y/Z |

Düz metin: OD-TÜ

Algoritmanın işleyişi aşağıdaki gibidir:

1-İkilideki harflerden ikisi de aynı satırda ise harflerin sağında yer alan harflerle yer değiştirirler.
2-İkilideki harflerden ikisi de aynı sütunda ise harflerin altlarında bulunan harfler ile değiştirilir.
3-İkilideki harfler aynı satırda veya aynı sütunda değilse, ilk harfi şifrelemek için bu harfin bulunduğu satır ve ikinci harfin bulunduğu sütunun kesişiminde bulunan harf alınır [6].

Şifreli metin: AC-PV

Anahtar sözcüğü bilinirse, aynı matrisi oluşturarak şifreleme algoritmasını tersten uygulayıp deşifreleme işlemini kolaylıkla yapılabilir. Örneğin; şifre metinin ilk ikilisi "AC" ikilisiydi. Aynı algoritma uygulanarak; "A"nın bulunduğu satırla "C"nin bulunduğu sütunun kesişimindeki harfin "O"; "A"nın bulunduğu sütunla "C"nin bulunduğu satırın kesişimindeki harfin ise "D" olduğu açıkça görülmektedir.

*2.12. Hill Şifresi (Hill Password)*

Leste S. Hill tarafından tasarlanan "Hill Şifresi" çok alfabeli şifreleme sistemlerine başka bir örnek olup çok alfabeli şifreleri daha pratik hale getirmesi bakımından oldukça önemlidir. Hill şifresi, tamamen lineer cebire dayandığı için kriptanalizi biraz teoriktir. Ancak yeterli düz metin ve onun karşılığı olan şifreli metin ele geçirildiği taktirde lineer cebir kullanılarak anahtar matris kolayca hesaplanabilir [11].

*2.13. Vernam Şifresi (Vernam Password)*

1917 yılında Amerikan Telefon ve Telgraf şirketinde çalışan Gilbert Vernam adındaki mühendis, yeni bir şifreleme tekniği geliştirdi. Vernam şifresiyle beraber, matematiğin kriptografide sistematik olarak kullanılmaya başlandığı görülmektedir. Vernam şifresini daha kolay anlayabilmek için işlemler bildiğimiz toplamaya göre yapılacaktır. Bunun için öncelikle alfabedeki her harf, aynı Hill şifresindeki gibi, alıcı ve gönderici arasındaki belli bir kurala göre belirlenmiş sayılarla eşleştirilir (Şekil 9) [7].

*2.14. Tek Seferlik Şifre (One-Time Password)*

Vernam şifresinin bulunmasından sonra 1917'de ABD ordusunda haberleşmede görevli binbaşı Joseph Mauborgne, Vernam şifresini tek seferde kullanmanın, yani her şifrelemeden sonra anahtarın değiştirilmesinin sistemi çok daha güvenli hale getireceğini keşfetmiştir. Bu şifreye ise tek seferlik şifre adı verilmiştir. Yıllar sonra (1940'larda) ünlü bilim adamı Claud Shannon çözülemez şifrelerin gerçekten mevcut olduğunu ve hatta bunların 30 yıldır bilindiğini göstermiştir. Bu bahsedilen şifre günümüzde de çözülemez olduğu ispatlanan "tek seferlik şifre"dir. Tek seferlik şifrenin güvenilirliği kesin olsa da ciddi bir dezavantajı vardır: anahtarın iletilme sorunu. Mesaj uzunluğunda bir anahtarın güvenli bir kanaldan iletilmesi gerekmektedir. Her seferinde de anahtar değişeceği için pratikte zor ve aynı zamanda masraflı bir sistemdir [11].

*2.15. Enigma (Enigma)*

2. Dünya Savaşı sırasında Nazi Almanyası tarafından gizlenmek istenen ve önemli olan mesajların şifrelenmesi ve tekrar çözülmesi amacı ile kullanılan şifre makinesidir. Enigma makinesi, ticari olarak 1920'li yılların başında kullanılmaya başlanmıştır [15]. Stratejik planların uygulanmasında kullanılan şifreleme ve deşifreleme teknikleri veya algoritmaları, buluşlar, şifre çözücü makineler bilgisayar biliminin gelişmesinde önemli derecede etkili olmuştur denilebilir [16].

| A | B | C | Ç | D | E | F | G | Ğ | H | I | İ | J | K | L | M | N | O | Ö | P | R | S | Ş | T | U | Ü | V | Y | Z |
|---|---|---|---|---|---|---|---|---|---|---|---|---|---|---|---|---|---|---|---|---|---|---|---|---|---|---|---|---|
| 0 | 1 | 2 | 3 | 4 | 5 | 6 | 7 | 8 | 9 | 10 | 11 | 12 | 13 | 14 | 15 | 16 | 17 | 18 | 19 | 20 | 21 | 22 | 23 | 24 | 25 | 26 | 27 | 28 |

Şekil 9. Vernam Şifresi tablosu (Vernam password table)

*2.16. Frekans Analizi (Frequency Analysis)*

Şifreli bir mesajı çözmek için, aynı dilde yazılmış yeteri derecede uzun bir metin bularak her bir harfin kullanım sıklığını hesaplamak gerekmektedir. Metinde en çok kullanılan harf, şifreli mesajda en çok kullanılan harfe denk gelmektedir. Aynı işlem, sırasıyla diğer harfler için de yapılmaktadır. Daha sonra mesajda bulunan harfler elde edilmiş olur. Al-Kindi, bu kriptanaliz yöntemine; bir dildeki her harfin bir kullanım sıklığı olmasından dolayı frekans analizi adını vermiştir. Bir harfin frekansı yeterince uzun metinler seçilerek bulunmaktadır. Frekans, o harfin metin içerisinde kaç kez kullanıldığının metindeki toplam harf sayısına bölünmesiyle bulunmaktadır. Bu sayı küçük sapmalar göstermekle birlikte, harflerin frekanslarının kendi aralarındaki büyüklük sıraları genellikle değişmemektedir [11,17]. Frekans analizi tekniği ile dile ait bazı özellikler kullanılarak şifreli metinden çözülmüş metnin elde edilmesi amaçlanmaktadır [18,19]. Şifreli metin içerisinde en sık kullanılan harf, metnin yazıldığı dilde en çok kullanılan harf ile eşleştirilir ve bu işlem tüm harfler için uygulanmaktadır. Her harfin kullanım sıklığı hesaplanmakta ve sırasıyla harflerin yerleştirilmesi devam etmektedir [20,21]. Aşağıda verilen örnek cümlede, A harfi için harf frekans hesaplaması anlatılmıştır.

Örnek: "Akif kasaba gitti ve et aldı."
Toplam harf sayısı: 23 ve A harfinin sayısı: 5
A harfinin frekansı: 5/23 = 0,21 olur.

Frekans analizi yönteminin uygulanabilmesi için uzun metinlere gereksinim duyulmaktadır. Çünkü uzun metinler, elde edilen verilerin güvenilir ve geçerli olmasını beraberinde getirmektedir [6,21,22]. Tüm bu şifreleme algoritmaları ışığında bu çalışmada, yeni bir şifreleme-deşifreleme algoritması geliştirilmiştir. Sozkonusu algoritmanın ayrıntıları üçüncü bölüme verilmiştir.

**3. TASARLANAN VERİ ŞİFRELEME ALGORİTMASI (DESIGNED DATA ENCRYPTION ALGORITHM)**

Çalışmada, yazılan bir metnin güvenli bir şekilde saklanması ve istenildiği takdirde kullanıcı tarafından tekrar çözülerek anlaşılır bir hal alması veya deşifre edilmesi amacıyla bir algoritma tasarlanmıştır. Bu kapsamda, uygulama Microsoft Visual Studio .NET C# paket programı kullanılarak geliştirilmiştir. Sözkonusu uygulamada Algoritma, "Tekli Alfabetik Yer Değiştirme" tekniğine göre tasarlanmıştır. Algoritmanın oluşturulmasında ikinci bölümde de anlatılmış olan Sezar Şifreleme Tekniği, Çoklu Alfabeli Algoritmaları ve Enigma'dan yararlanılmıştır. Bu algoritmaların kullanılma sebebi, sözkonusu algoritmalarla şifrelenen metinlerin çözülmesinin daha zor olması ve daha hızlı performansa sahip olmasıdır. Böylece çözülmesi güç bir algoritma geliştirmek amaçlanmıştır. Veri Şifreleme Algoritmasında kelimenin içinde bulunan harfler indislerine göre; yani harfin sırasına göre gruplanarak tek ya da çift sayı olup olmadığına bakılmak suretiyle, sonuca göre farklı iki şifreleme grubundan geçirilmiştir. Eğer harfin indis sırası tek (1-3-5-7-9-...) ise, ilk şifreleme grubundan; indis sırası çift (2-4-6-8-10-...) ise ikinci şifreleme grubundan geçirilmektedir. Her bir şifreleme grubu üç ayrı, toplamda altı şifre alfabesinden oluşmaktadır. Harfin hangi şifreleme grubuna gireceği belirlendikten sonra girdiği şifreleme grubunda eldeki harf düz alfabe grubundan bulunarak şifre alfabesinde karşılık gelen harf alınmaktadır. Şifre alfabesinde bulunan harf alınarak ikinci düz alfabeye bakılmaktadır. Düz alfabede, bir önceki şifre alfabesinden elde edilen harf bulunarak karşılık olarak gelen şifre alfabesindeki karşılığı alınmaktadır. Üçüncü adımda da; bir önceki şifre alfabesinden elde edilen harf alınarak, üçüncü düz alfabede bulunmaktadır. Daha sonra karşılık olarak gelen şifre alfabesindeki karşılığı alınmaktadır. Elde edilen harf son şifre alfabesinde bulunmakta olup şifre alfabesinde kendinden bir önceki harfle yer değiştirmektedir. Elde bulunan metindeki her harf bu işlemlerden tek tek geçirilerek şifreli metin elde edilmiş olmaktadır. Son şifre alfabesinde, bulunan harfi kendinden önceki harfle yer değiştirerek elde edilecek şifreli metni biraz daha karmaşık bir yapıya sahip olması amaçlanmıştır. Burada kullanılan yöntem, "Sezar Şifreleme" algoritmasına benzemekle birlikte alfabedeki harflerin rastgele yerleştirilmesi ve kendinden önceki harfle yer değiştirmesiyle bu yöntemden ayrılmaktadır.

Veri Şifreleme Algoritmasında iki ayrı şifreleme grubu kullanmak, harfleri indislerine göre ayırarak şifrelemek ve Frekans Sıklığı Yöntemi kullanmak suretiyle şifrelenen metinlerin çözülmesinin önüne geçilmeye çalışılmıştır. Harflerin indis numaralarının farklı olduğu kelimelerde aynı harf farklı harflerle şifreleneceğinden frekans sıklığı yöntemi pek işe yaramamakta ve farklı sonuçlar ortaya çıkmaktadır. Tasarlanan algoritmada iki farklı şifreleme grubunun bulunması, aynı harfin iki farklı harf olarak elde edilmesi demektir. Örneğin, A harfi bir şifreleme grubundan V harfi olarak çıkarken, diğer şifreleme grubundan Ğ harfi olarak çıkacaktır. Bu da Frekans Analizi Yöntemini kullanarak şifrenin çözülme ihtimalini oldukça düşürecektir. Çünkü, A harfi şifreli metinde iki farklı harfle temsil edilmektedir. Birden fazla şifre alfabesinin kullanılması ve her şifre alfabesindeki harflerin belli bir düzen olmaksızın rastgele yerleştirilmesi, şifreli metnin çözümünü zorlaştıran unsurlardır. Çünkü, şifrenin çözülme çabası içerisinde her alfabe için 29! adet farklı alfabe denenmesi ve bu kadar denemenin 7 ayrı şifre alfabesini bulmak için yapılması gerekmektedir. Bu da zaman alıcı, zor ve zahmetli bir iştir. Şekil 10'da, Veri Şifreleme Algoritması'nın akış diyagramı verilmiştir.

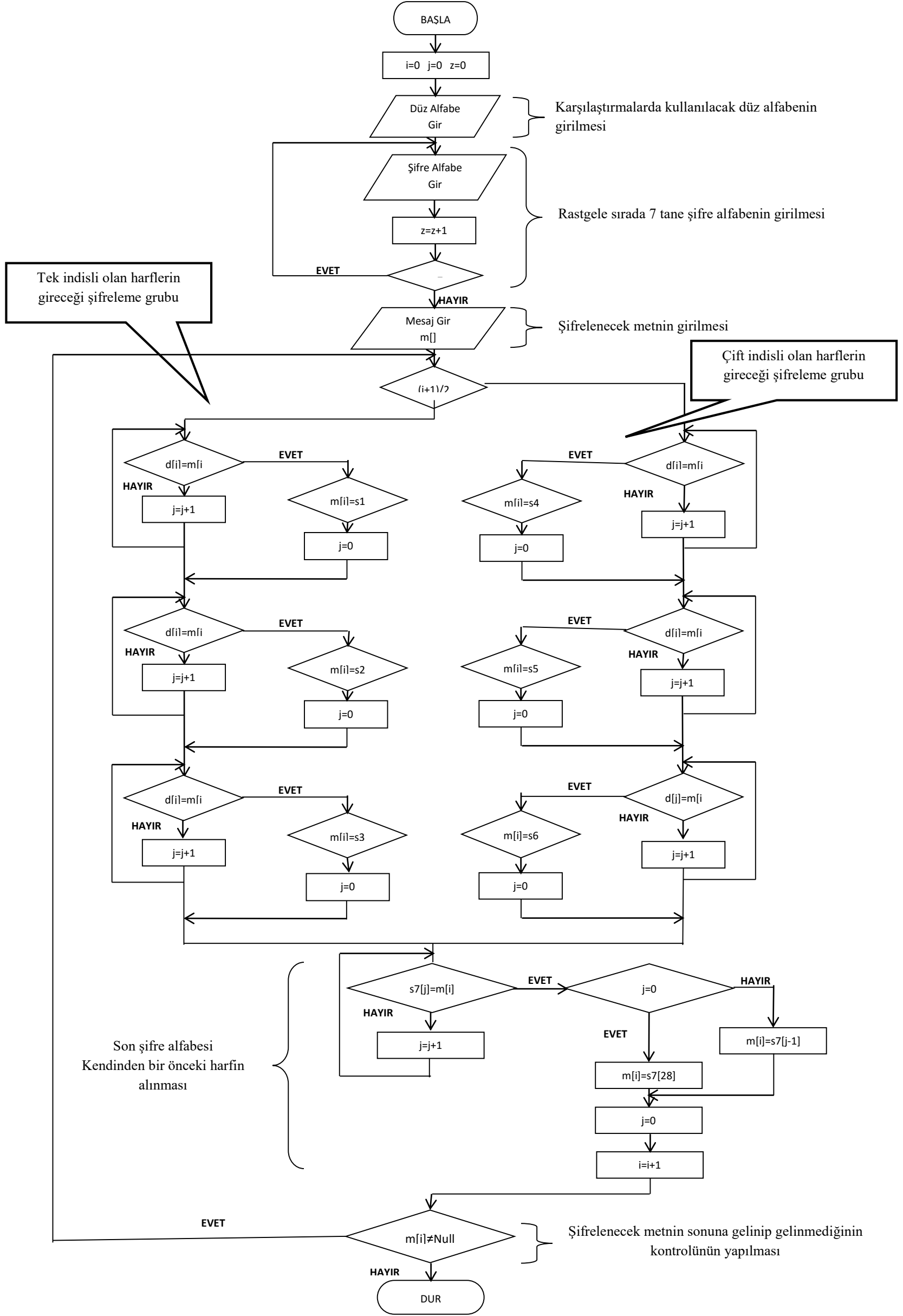

Şekil 10. Veri şifreleme algoritması akış diyagramı (Data encryption algorithm flow chart)

*3.1. Şifreleme ve Deşifreleme Adımları (Encryption and Decryption Steps)*

Çalışma kapsamında geliştirilen Şifreleme ve Deşifreleme algoritmalarının adımları şu şekilde gerçekleşmektedir.

Şifreleme Algoritmasının Adımları

- Girilen mesaj bir diziye (m[i]) aktarılır.

- Sonrasında bir döngü başlar ve bu döngü dizinin son karakteri Null (boş) olana kadar devam eder.

- İlk adımda "(i+1)%2=0" işlemi ile o anda sorgulanan indisin tek ya da çift bir sayı olup olmadığı sorgulanır. Eğer işlemin sonucu "0" ise bu durumda indis çift sayı olur ve 2. Şifreleme Grubunda işlem görür, sonuç "0" değilse bu durumda 1. Şifreleme Grubunda işlem görür.*(Örnek olarak "Örnek Şifreleme Grupları"ndan 1. Şifreleme Grubu seçilmiştir.)*

Hangi şifreleme grubunda işlem göreceği belirlendikten sonra;

- Şifreleme grubunun ilk sırasındaki bölümde şifrelenecek harfin düz alfabedeki eşiti bir döngü yardımıyla bulunur. Düz alfabede eldeki harfin eşiti bulunduğunda bu harfin indis numarasına denk gelen şifre alfabedeki harf karşılığı ile değiştirilir*(m[i]=s1[j])*.

- Değiştirme tamamlandığında ikinci sıradaki bölüme geçilir. Burada da bir önceki adımda elde edilen harfin düz alfabedeki eşiti döngü yardımıyla bulunur. Düz alfabede eldeki harfin eşiti bulunduğunda bu harfin indis numarasına denk gelen şifre alfabedeki harf karşılığı ile değiştirilir*(m[i]=s2[j])*.

- Değiştirme tamamlandığında üçüncü sıradaki bölüme geçilir. Burada da bir önceki adımda elde edilen harfin düz alfabedeki eşiti döngü yardımıyla bulunur. Düz alfabede eldeki harfin eşiti bulunduğunda bu harfin indis numarasına denk gelen şifre alfabedeki harf karşılığı ile değiştirilir*(m[i]=s3[j])*.

- Bu üç bölümde elde edilen harf son şifre alfabeye gelir. Burada bir döngü yardımıyla eldeki harfin şifre alfabedeki eşiti bulunur. Bulunan harf, şifre alfabede kendinden önce gelen harfle yer değiştirir*(m[i]=s7[j])*.

Bütün bu işlemlerin sonunda elde edilen harf, şifrelenmek üzere işleme giren harfin şifrelenmiş karşılığı olur.

Daha sonra sözkonusu şifrelenmiş metin istenildiği durumda deşifre edilebilmektedir. Bu işlem ise şu adımlar sonucu gerçekleşmektedir.

Deşifreleme Algoritmasının Adımları

- Metin bir diziye aktarılır.

- Harfin bulunduğu indise göre hangi şifreleme grubuna gireceğine karar verilir. *(Örnek olarak "Örnek Şifreleme Grupları"ndan 1. Şifreleme Grubu seçilmiştir.)*

- Elde bulunan şifreli harf son şifre alfabesinde bulunur, kendinden sonraki harf ile yer değiştirir.

- Değiştirme tamamlandığında üçüncü sıradaki bölüme geçilir. Burada da bir önceki adımda elde edilen harfin şifre alfabesindeki eşiti döngü yardımıyla bulunur. Şifre alfabesinde eldeki harfin eşiti bulunduğunda bu harfin indis numarasına denk gelen düz alfabedeki harf karşılığı ile değiştirilir.

- Değiştirme tamamlandığında ikinci sıradaki bölüme geçilir. Burada da bir önceki adımda elde edilen harfin şifre alfabesindeki eşiti döngü yardımıyla bulunur. Şifre alfabesinde eldeki harfin eşiti bulunduğunda bu harfin indis numarasına denk gelen düz alfabedeki harf karşılığı ile değiştirilir.

- Değiştirme tamamlandığında şifreleme grubunun ilk sırasındaki bölümde eldeki şifrelenecek harfin şifre alfabesindeki eşiti bir döngü yardımıyla bulunur. Şifre alfabesinde eldeki harfin eşiti bulunduğunda bu harfin indis numarasına denk gelen düz alfabedeki harf karşılığı ile değiştirilir. Bütün bu işlemlerin sonunda elde edilen harf, şifresi çözülmek üzere işleme giren harfin şifresinin çözülmüş karşılığı olur.

Geliştirilen algoritmanın şifreleme-deşifreleme adımları örnek bir metin üzerinden aşağıda anlatılmıştır.

*Şifrelenecek Örnek Metin*

> Mikroişlemci, komutları sırasıyla işleyerek sonuç elde eden karmaşık bir bilgisayar birimidir. İşlemci tarafından işlenecek komut satırları küme halinde belleğin kod bölümünde bulunmaktadır. Bu komutlar sırasıyla bazen de sırasız olarak alınıp işlemciye getirilir ve bir dizi işlemden sonra sıradaki komutun aynı yolla işlenmesine geçilir. Birinci sırada işlenen komutun sonucu ikinci sırada işlenen komutu bazen ilgilendirir bazen de ilgilendirmez. Bütün bu işlemler bir grup birim içerisinden geçirilerek gerçekleşir. İşlenen bu komutun bir sıra halinde işlemden geçirilmesi merhalesine iş hattı tekniği denir.

*Örnek Şifreleme Grupları*

**1. ŞİFRELEME GRUBU**

| DÜZ | A | B | C | Ç | D | E | F | G | Ğ | H | I | İ | J | K | L | M | N | O | Ö | P | R | S | Ş | T | U | Ü | V | Y | Z |
|---|---|---|---|---|---|---|---|---|---|---|---|---|---|---|---|---|---|---|---|---|---|---|---|---|---|---|---|---|---|
| ŞİFRE | B | S | Y | K | A | D | M | R | Ş | Ç | O | Z | E | N | C | G | H | İ | F | İ | L | Ğ | Ö | V | P | T | U | Ü | J |
| DÜZ | A | B | C | Ç | D | E | F | G | Ğ | H | I | İ | J | K | L | M | N | O | Ö | P | R | S | Ş | T | U | Ü | V | Y | Z |
| ŞİFRE | A | Z | C | G | H | J | N | B | Ö | Ç | L | Ş | Ğ | Ü | İ | P | I | K | T | Y | R | E | V | D | F | S | U | O | M |
| DÜZ | A | B | C | Ç | D | E | F | G | Ğ | H | I | İ | J | K | L | M | N | O | Ö | P | R | S | Ş | T | U | Ü | V | Y | Z |
| ŞİFRE | P | I | V | K | Z | C | H | N | G | S | U | D | A | F | İ | R | E | Ü | J | Ğ | Ş | L | T | Y | B | Ö | Ç | O | M |

**2. ŞİFRELEME GRUBU**

| DÜZ | A | B | C | Ç | D | E | F | G | Ğ | H | I | İ | J | K | L | M | N | O | Ö | P | R | S | Ş | T | U | Ü | V | Y | Z |
|---|---|---|---|---|---|---|---|---|---|---|---|---|---|---|---|---|---|---|---|---|---|---|---|---|---|---|---|---|---|
| ŞİFRE | S | A | Ş | Z | R | Ö | Ç | E | İ | J | K | T | Y | O | N | P | B | M | H | Ü | D | V | L | U | I | G | C | F | Ğ |
| DÜZ | A | B | C | Ç | D | E | F | G | Ğ | H | I | İ | J | K | L | M | N | O | Ö | P | R | S | Ş | T | U | Ü | V | Y | Z |
| ŞİFRE | Ş | V | H | Ö | Ç | D | A | J | L | İ | R | E | P | I | Z | C | F | N | Ğ | Ü | K | T | Y | B | G | S | U | O | M |
| DÜZ | A | B | C | Ç | D | E | F | G | Ğ | H | I | İ | J | K | L | M | N | O | Ö | P | R | S | Ş | T | U | Ü | V | Y | Z |
| ŞİFRE | Z | Ş | N | I | D | Y | S | M | H | Ç | V | R | L | Ğ | C | Ü | P | K | G | B | U | Ö | J | F | A | T | İ | O | E |

**SON ŞİFRELİ ALFABE**

| D | Ö | J | A | S | Z | B | N | Ü | L | C | R | Ş | E | Ç | Y | Ğ | F | I | T | H | G | İ | O | K | V | M | P | U |
|---|---|---|---|---|---|---|---|---|---|---|---|---|---|---|---|---|---|---|---|---|---|---|---|---|---|---|---|---|

*Şifrelenmiş Örnek Metin*

Pbfpzavfğ öküğöküav cşktütutp öbopd aamt ayag pbfdktpj lru lrkucjğsğf lrurfrmru. Cşktfic zğuğykoyvg cşktotntp pbfpz öğzkuavfğ püft bğkroya ltkaaçcg pbm lcküfüoya lpkpoıvmzğmğu. Li pbfpzavf öküğöküav lğhto mt öküğökh gavfvm vağğğd cşktııcsa rtzrurkru tt lru mrhr cşktfyag öbofv öküğmğpr pbfpzpo vsok übkav cşktoıajcga rtdrkru. Lruroic öküğmğ cşktoto pbfpzpo öbopnp cmcgnr öküğmğ cşktoto pbfpzp lğhto carrktoycfcf lğhto mt carrktoycffth. Lüzüo li cşktfaaf lru rfid lrurf cnafcjcgmto rtdrurktutp rtunamktçru. Cşktoto li pbfpzpo bru öküğ bğkroya cşktfyag radrurkıaj ftuzvaajcga cş bğzvğ ztpgccc mtorkru.

*Şifrelenmiş Örnek Metnin Çözümü*

Şifreleme algoritmasına göre şifrelenmiş bir metnin şifresinin çözülebilmesi için şifrelemede kullanılan bütün şifre alfabelerin aynı sırada metni çözecek kişinin elinde olması ve aynı zamanda çift veya tek indisler için hangi şifreleme gruplarının kullanıldığının bilinmesi gereklidir.

Eldeki şifre alfabelerine göre çözüm yapabilmek için;

- Metin bir diziye aktarılır.

- Harfin bulunduğu indise göre hangi şifreleme grubuna gireceğine karar verilir. *(Örnek olarak "Örnek Şifreleme Grupları"ndan 1. Şifreleme Grubu seçilmiştir.)*

- Elde bulunan şifreli harf son şifre alfabesinde bulunur ve kendinden sonraki harf ile yer değiştirir.

- Değiştirme tamamlanınca üçüncü sıradaki bölüme geçilir. Burada da bir önceki adımda elde edilen harfin şifre alfabesindeki eşiti döngü yardımıyla bulunur. Şifre alfabesinde eldeki harfin eşiti bulununca bu harfin indis numarasına denk gelen düz alfabedeki harf karşılığı ile değiştirilir.

- Değiştirme tamamlanınca ikinci sıradaki bölüme geçilir. Burada da bir önceki adımda elde edilen harfin şifre alfabesindeki eşiti döngü yardımıyla bulunur. Şifre alfabesinde eldeki harfin eşiti bulununca bu harfin indis numarasına denk gelen düz alfabedeki harf karşılığı ile değiştirilir.

- Değiştirme tamamlanınca şifreleme grubunun ilk sırasındaki bölümde eldeki şifrelenecek harfin şifre alfabesindeki eşiti bir döngü yardımıyla bulunur. Şifre alfabesinde eldeki harfin eşiti bulununca bu harfin indis numarasına denk gelen düz alfabedeki harf karşılığı ile değiştirilir.

Bütün bu işlemlerin sonunda elde edilen harf, şifresi çözülmek üzere işleme giren harfin şifresinin çözülmüş karşılığı olur.

## 4. SONUÇ VE ÖNERİLER (CONCLUSIONS AND SUGGESTIONS)

Bu çalışmada, Sezar Şifreleme Tekniğinden, Çoklu Alfabeli Algoritmalardan ve Enigma'dan yararlanılarak yeni bir şifreleme-deşifreleme algoritması geliştirilmiştir. Tasarlanan veri şifreleme algoritmasında harflerin yerleri belirli bir kurala göre yer değiştirmiştir. Bu yüzden metinde bulunan harfler indis numaralarına göre iki farklı şifreleme grubundan geçirilmiştir. Her şifreleme grubunun içinde de üç farklı şifre alfabeden geçirilip elde edilen şifreli harf sonuncu şifre alfabesinde kendinden bir önceki harf alınarak şifrelenmiştir. Burada iki farklı şifreleme grubu kullanılmasının amacı, farklı indis sırasında bulunan aynı harfi şifreleme sonucunda iki farklı harf olarak elde etmektir. Böylece şifrenin kırılması daha zor bir hal alacaktır.

Veri şifreleme algoritması, tek rakamlı indislerin ilk şifreleme grubunda, çift rakamlı indislerin ikinci şifreleme grubunda şifrelenmesi ve her iki şifreleme grubundan çıkan harfin sonuncu şifre alfabede kendinden önceki harfle yer değiştirmesi esasına dayanmaktadır. Bu tekniğin farklı yöntemlerle geliştirilmesi de mümkündür. Örneğin; tek rakam veya çift rakamlı indisler olarak ayırdıktan sonra, harfleri tek tek şifrelemek yerine, belirli uzunluklardaki bloklara bölerek ve bloklar farklı şifreleme gruplarından geçirilerek şifreleme işlemi yapılabilir. Şifreleme grubu sayısı artırılabilir. Her iki şifreleme grubu için ortak kullanılan son şifre alfabeden farklı bir tane daha eklenerek her şifreleme grubunda farklı bir şifre alfabe olması sağlanabilir. Son şifre alfabeden geçirilerek elde edilen şifrelenmiş harflere sayısal değerler verilerek saklanması sağlanabilir.

Geliştirilen model ile girilen bir metnin şifrelenmesi ve tekrar deşifre edilmesi işlemleri hatasız bir şekilde yapılabilmiş ve başarılı sonuçlar elde edilmiştir.

## KAYNAKLAR (REFERENCES)

[1] T. Yerlikaya, E. Buluş, ve N. Buluş, "Kripto algoritmalarının gelişimi ve önemi", Akademik Bilişim Konferansları 2006-AB2006, Denizli-Türkiye, Şubat-2006.

[2] M., Baykara, R., Daş, ve İ. Karadoğan, (2013, May). Bilgi Güvenliği Sistemlerinde Kullanılan Araçların İncelenmesi. In 1st International Symposium on Digital Forensics and Security (pp. 231-239).

[3] I., Cisa Review Manual 2009, Isaca Press, Rolling Meadows, 2009.

[4] M. Gülmüş, "Kurumsal Bilgi Güvenliği Yönetim Sistemleri ve Güvenliği", Yıldız Teknik Üniversitesi Fen Bilimleri Enstitüsü Yüksek Lisans Tezi, 2010.

[5] G. Canbek, Ş. Sağıroğlu, "Bilgi, Bilgi Güvenliği ve Süreçleri Üzerine Bir İnceleme", Politeknik Dergisi, 9(3):69-72.

[6] C., Çimen, S., Akyelek, E., Akyıldız, "Şifrelerin Matematiği: Kriptografi", ODTÜ Yayınları, 2009.

[7] Kumari, S. (2013). *Recurrent Sequences and Cryptography* (Doctoral dissertation, National Institute of Technology Rourkela, India).

[8] A., Jones, "Information Warfare-what has been happening?", ComputerFraud& Security, 4-7 (November 2005).

[9] A. Salomaa, "Public-Key Cryptography", Springer-Verlag, New York, 1990

[10] S., Keren, "Şifreleme Bilimi Üzerine", Lisans Tezi, Yıldız Teknik Üniversitesi, İstanbul, 2011.

[11] İnternet: Vikipedi, "Ebced hesabı", http://tr.wikipedia.org/wiki/Ebced_hesabı (07.04.2016).

[12] B., Kuru, "Zigzag Şifrelemesi", http://www.bilgisayarkavramlari.com/2009/06/01/zigzag-sifrelemesi-zigzag-cipher/ (05.05.2016)

[13] B., Keskin, "Bilişim Sistemlerinin Stratejik Yönetim Açısından Önemi İçin Kriptografi", Yüksek Lisans Tezi, Gebze İleri teknoloji Enstitüsü, Kocaeli, 2010

[14] Karahoca, A., & Sarısakal, M. N. (2011). Des Algoritmasını Kullanan Güvenilir Bir E-Posta İletim Uygulaması: Tuğra. *Iu-Journal Of Electrical & Electronics Engineering*, *1*(1).

[15] T., Yerlikaya, E., Buluş, N., Buluş, "Kripto Algoritmalarının Gelişimi ve Önemi", Lisans Tezi, Trakya Üniversitesi Bilgisayar Mühendisliği Bölümü, Edirne, 2011.

[16] A., Eskici, (2004). Polybius Şifresi. Matematik Dünyası Dergisi.

[17] A., Coşkun, ve Ü. Ülker, (2013). Ulusal Bilgi Güvenliğine Yönelik Bir Kriptografi Algoritması Geliştirilmesi ve Harf Frekans Analizinine Karşı Güvenlirlik Tespiti. International Journal of Informatics Technologies, 6(2), 31.

[18] D. Arda, E. Buluş, "Türk Alfabesi ve Yapısal Özellikleri Kullanılarak Tek Alfabeli Yerine Koymada Şifreleme ve Kriptanaliz", 20. Türkiye Bilişim Kurultayı, İstanbul, 2003.

[19] D,. Arda, E. Buluş, T. Yerlikaya, "Türkiye Türkçesi'nin Bazı Dil Karakteristik Ölçütlerini Kullanarak Vigenere Şifresi ile Şifreleme ve Kriptanaliz", ELECO'2004 Elektrik-Elektronik-Bilgisayar Mühendisliği Sempozyumu ve Fuarı, Bursa, 2004.

[20] S., Soyalıç, "Kriptografik Hash Fonksiyonları ve Uygulamaları", Yüksek Lisans Tezi, Erciyes Üniversitesi Fen Bilimleri Enstitüsü Matematik Anabilim Dalı, Kayseri 2005.

[21] H. N., Buluş, "Temel Şifreleme Algoritmaları ve Kriptanalizlerinin incelenmesi", Yüksek Lisans Tezi, Trakya Üniversitesi Fen Bilimleri Enstitüsü, Edirne 2006.

[22] M. Ü., Çeşmeci, "Kriptoloji Tarihi", UEKAE Dergisi, 2009.